# An Undergraduate Course on the Statistical Principles of Research Study Design

Lee Kennedy-Shaffer, PhD

*Department of Biostatistics, Yale School of Public Health, New Haven, CT, USA*

Correspondence: Lee.kennedy-shaffer@yale.edu

# An Undergraduate Course on the Statistical Principles of Research Study Design

The undergraduate curriculum in statistics and data science is undergoing changes to accommodate new methods, newly interested students, and the changing role of statistics in society. Because of this, it is more important than ever that students understand the role of study design and how to formulate meaningful scientific and statistical research questions. While the traditional Design of Experiments course is still extremely valuable for students heading to industry and research careers, a broader study design course that incorporates survey sampling, observational studies, and the basics of causal inference with randomized experiment design is particularly useful for students with a wide range of applied interests. Here, I describe such a course at a small liberal arts college, along with ways to adapt it to meet different student and instructor background and interests. The course serves as a valuable bridge to advanced statistical coursework, meets key statistical literacy and communication learning goals, and can be tailored to the desired level of computational and mathematical fluency. Through reading, discussing, and critiquing actual published research studies, students learn that statistics is a living discipline with real consequences and become better consumers and producers of scientific research and data-driven insights.

**Introduction**

The discussion of statistics and data science often surrounds methods for the analysis of data sets. Because of this, undergraduate institutions generally structure their curricula around a series of courses teaching successively more advanced methods to students, incorporating applications throughout. This paradigm has been challenged lately, as educators have placed renewed emphasis on conceptual understanding and communication, as well as ethics and data science skills (Horton and Hardin 2015). Similarly, the relationship between scientific research (here construed broadly to include any field approaching hypotheses using the scientific method) questions, study design, and statistical analysis deserves greater emphasis in the undergraduate curriculum. Box (1990,p. 251) noted that "Statistics is, or should be, about scientific investigation and how to do it better" and Higgins (1999,p. 2) reiterated this, noting that "the nonmathematical aspects of the undergraduate discipline are of significant benefit to society."

Study design has long played a key role in the development of statistics and in the role of statisticians working in various research areas. In the early history of statistics departments, this often focused on experimental design with an agricultural or industrial focus, while biostatistics departments focused on the design and analysis of clinical trials (see Agresti and Meng 2013). Textbooks specialized as well, with experimental design and survey sampling becoming specialized subfields of statistics (Agresti 2023). In most cases, the focus in statistical courses tended to be on conducting the analyses for various designs (Higgins 1999). These Design and Analysis of Experiments (DoE) courses have often persisted, in both statistics and engineering departments, as the main "design" course offering available (Vazquez and Xuan 2024).

The statistics graduate of today, however, enjoys a wider range of options for careers and post-graduate education, and so is likely to need a broader understanding of the use of statistics in study design. Moreover, the ubiquity of statistics and data analysis in the modern world demands that college graduates be equipped with statistical literacy that cuts across applied areas and embraces the principles behind making meaning from data (Lerner and Gelman 2024) as part of a broad updated undergraduate statistics education (Horton and Hardin 2015). The need for such courses has been noted by a variety of authors. Higgins (1999) suggested "nonmathematical" statistics courses covering topics such as "Planning and Managing Surveys", "Planning and Managing Scientific Experiments", and "Communicating Statistical Ideas." Utts (2003,p. 74) called for emphasis on "how to integrate information from study design to final conclusions in a meaningful way." And Brearley, Rott, and Le (2023) described the need for and value of teaching statistical literacy at the graduate and professional level.

To engage with the world and participate in scientific and data-driven conversations, literacy around study designs and the ability to interpret and communicate their benefits and drawbacks are key. For "educated citizens", writes Utts (2003,p. 78): "What good is it to know how to carry out a *t* test if a student can not read a newspaper article and determine that hypothesis testing has been misused?" The demand for statistics training to incorporate the key role of statistician as communicator or bridge between disciplines and between subject-matter knowledge and data has been prominent for decades (see, e.g., Higgins 1999; Horton and Hardin 2015). This has become only more important in the data science era (see, e.g., American Statistical Association Undergraduate Guidelines Workgroup 2014; Hardin et al. 2015). These topics serve crucial roles for students going into business, technology, natural and social sciences, policy, or many other fields, and provide fundamental background, conceptual understanding, and a grasp of the role of the statistician to those pursuing further statistics and data science work (Utts 2003).

This manuscript presents the author's approach to a modern statistical study design course at the upper undergraduate level. This semester-long course has been offered from 2020–2024 at Vassar College, a selective liberal arts college in the United States. It was an elective available in the mathematics and statistics major and was taken primarily—though not exclusively—by students either majoring or minoring in mathematics and statistics. Many students had a non-mathematics major, either as their primary or double major, mostly in either the natural or social sciences. Because of this, the course sought to be relevant to a wide variety of applied fields and draw applications and examples from these areas. Because of the liberal arts setting, it also sought to build a humanistic understanding of statistical and scientific principles and the role of statistical evidence in a pluralistic society.

The next section presents an overview of the course as taught by the author, encompassing the learning objectives, course topics, course format and assignments, and primary and secondary literature used as applications and examples. Appendix A contains relevant sections of the syllabus and the full reading list. The third section discusses the broader role of this course in the undergraduate curriculum and its relationship to statistical education guidelines. The fourth section discusses challenges and opportunities in such a course offering, and the final section presents concluding remarks.

**Course Overview**

*Learning Objectives*

The learning objectives, shown in Box 1, focus primarily on conceptual understanding, identification of key statistical principles, and communication. The final two learning objectives connect more specifically to the mathematical and statistical methodology presented. For similar courses with more computational or proof-based mathematical components, relevant learning objectives can be adapted or added.

> Box 1. Learning objectives used in the course.
>
> By the end of the course, students will be able to:
>
> 1. identify the key statistical and scientific goals of research studies;
> 2. explain the benefits, drawbacks, and limitations of various study designs and features;
> 3. read methods sections of scientific papers, understand the statistical concepts discussed therein, and assess the choices made in the study design;
> 4. understand the role of statisticians in the design and execution of studies;
> 5. communicate the course concepts to audiences with varying mathematical and statistical backgrounds;
> 6. use the mathematical tools of probability and statistics to evaluate study design features; and
> 7. explain how scientific goals affect statistical needs and how statistical limitations of study designs shape feasible scientific goals.

These objectives differ from those of many statistics courses, which may focus on specific topics or models. However, they build on recommendations of the GAISE report for introductory course learning outcomes, especially the first three: "teach statistical thinking", "focus on conceptual understanding", and "integrate real data with context and purpose" (GAISE College Report ASA Revision Committee 2016). In this way, the course fits into a broader statistics curriculum (American Statistical Association Undergraduate Guidelines Workgroup 2014; see below for more discussion) and complements or acts as a bridge to more topic- or methods-focused courses.

*Course Topics*

The course framework permits a high degree of modularity and adaptation to the backgrounds and interests of the instructor and students. In general, I have divided it into four broad units: sampling, basic randomized experiments, advanced concepts in randomized experiments, and observational studies. Appendix A contains relevant sections of the syllabus and schedule with more details.

The sampling unit focuses on the statistical principles of variance and bias, both conceptually and mathematically. This reinforces and expands upon core statistical concepts such as variance, expectation, and estimation and inference through examples based on physical sampling—using sampling as a frame for repeated scientific measurement—and survey sampling such as political polling. The study design features of stratified and cluster sampling are introduced, as are the statistical concepts of sample size and mean squared error; all of these will be expanded upon in the second unit.

The unit introducing randomized experiments brings in causal ideas such as exchangeability, consistency, and positivity, as well as the idea of statistical power in hypothesis testing. Ethics of research also form a core part of this unit, demonstrating the link between the role of science in society, ethical frameworks, and the feasibility and statistical properties of studies. Case studies in this unit have included major clinical trials such as those of the weight-loss and diabetes drug semaglutide (see "Introduction to Randomized Experiments: Semaglutide Studies" below), COVID-19 treatment and vaccine studies, as well as experiments in cognitive science, economics, labor markets, and housing mobility. In addition, historical examples with key ethics lessons are discussed, including the Tuskegee syphilis study (Jones 1993; Reverby 2000, 2012).

Advanced randomized experiments topics can include many of those frequently taught in more traditional DoE courses (Smucker et al. 2023; Vazquez and Xuan 2024). Key statistical principles and concepts discussed include the bias-variance tradeoff, statistical efficiency, and estimands and effect sizes. The instructor can introduce these through designs and methods such as stratified/blocked or matched-pairs randomization, multi-arm and factorial designs, and cluster randomized trials. While the depth of coverage of these topics does not match that in standard DoE courses, the breadth can spur student thinking in those areas while maintaining applicability across applied fields. Case studies in these areas have included experiments run in political science, agriculture and ecology, public health, and education, along with large-scale A/B testing in technology and business settings.

Finally, the unit on observational studies discusses the ideas of causal inference when randomization is infeasible. Causal notions such as exchangeability, consistency, internal and external validity, transportability and generalizability shape the topics taught, with implications for the sampling and randomized designs discussed earlier. Specific designs and fields of application presented have included large-scale cohort studies in nutrition and health, quasi-experimental study designs used in economics, political science, and policy analysis (see"Quasi-experimental Designs: Economics and Policy Evaluation" below), and case-control studies used in legal controversies.

The course concludes by inviting students to consider the role of quantitative research studies in society. With their newfound statistical understanding of the challenges of designing, conducting, and interpreting these studies, students can engage with questions of understanding, certainty, and communication. They can also engage with questions about the limitations of the statistical approach to knowing or interpreting a complicated world (Messeri and Crockett 2024; Stevenson 2023).

*Course Format and Assignments*

This course was taught in twice-weekly, 75-minute sessions, a standard format at the institution. Generally, the first session in each week was devoted to introducing the concepts, terminology, and methods of that week's topic, with an example study (real or hypothetical) guiding and making concrete the terms used (see the Schedule in Appendix A.2). The content was delivered partially in a lecture style, with active learning opportunities frequently sprinkled throughout. In a small class, this can include full participation from the students; in larger classes, pair, small group, or response poll activities with brief discussion can be used instead. Active learning opportunities include students brainstorming ways to answer certain questions, how to iteratively refine designs and questions, and ways to mitigate statistical challenges.

The second session each week was devoted to guided discussion of assigned readings. The readings would include one or two published scientific studies, sometimes supplemented with news articles or commentary on the studies. If multiple studies were used, they would be within the same general field of application. The studies were selected to (in rough order of priority): exemplify the designs and statistical features presented; clearly describe the methods and design used; relate to student interests and ensure a variety of application domains; and be timely or of general interest. Students particularly appreciate variety or papers that relate to their primary field of interest, as reflected in evaluations ("Course Evaluations" 2023; "End of Course Survey" 2024). The studies and optional readings from the most recent version of the course are included in Appendix A.3. The instructor created a Study Reading Guide (Appendix A.4) for students to use for note-taking while reading and to self-assess understanding. This guide was presented by the instructor for a case study in week two and used optionally thereafter; it was never collected or graded. Discussion would generally occur in small groups, guided by instructor questions presented with time constraints. Reporting out from small groups to the full class would occur where appropriate, including filling out particularly challenging aspects of the Study Reading Guide as a group.

Assessment came from class participation and engagement in discussions, five problem sets assigned throughout the semester, an individual midterm project, and a group final project (see Appendix A.1). The problem sets assess student progress across the learning objectives through a mix of mathematical engagement with statistical properties, computational analysis and simulation, and conceptual understanding and communication. For example, one problem set has students compute power and required sample size using built-in R functions, conduct simulations to assess the sensitivity of power to assumptions, use probability and mathematical tools to compute bias and variance for estimands, and discuss the ethics of a historical study and the statistical challenges that must be met to conduct the study ethically.

The midterm project invites students to assess a published study and communicate their understanding. Students are given a small list of articles to choose from. For their chosen study, they create two assessments: a statistical critique that analyzes the study as if they were the statistical reviewer, specifically commenting on the design and analysis choices; and a public-facing piece (e.g., newspaper op-ed, infographic, or podcast episode) that critically summarizes the results of the study and communicates the conclusions and limitations of the study. This specifically builds students' communication skills, asking them to consider both a statistically advanced audience and a lay audience.

For the final project, small groups of students design a hypothetical study. They undertake the full study design process, choosing a scientific research question, population of interest, outcomes and variables to be measured, statistical analysis method and estimator, and determine the appropriate sample size. The students then present the design to the class and receive feedback from fellow students and the instructor. The final submission is a study design proposal (akin to a grant submission or a study protocol article).

Instructors familiar with DoE courses will note that, unlike for many such courses, this course did not include a full analysis of a conducted experiment, either real or simulated (Smucker et al. 2023; Vazquez and Xuan 2024). As analysis methods are not a major part of this course curriculum, this element is de-emphasized. The hypothetical study design project also allows students to consider any field of application and design type, connecting it more closely to their disciplinary interests. This provides a different approach to meet similar learning goals as the conducted experiment (Hunter 1977). However, it has the downside of not exposing students to the challenges of the actual implementation, measurement, and analysis of data. Instructors may determine which type of project best fits their goals.

**Example Topics**

To illustrate the connection between the study design concepts, mathematical and computational topics, and the assigned studies, this section describes the content from two weeks of the course, with special emphasis on the aspects of the assigned studies that are discussed.

*Introduction to Randomized Experiments: Semaglutide Studies*

In the first week of the randomized experiments unit, the principles of randomization are introduced through clinical trials. The principles of exchangeability, consistency, and positivity are introduced, with or without formal counterfactual or potential outcome notation. The introduction to the topic can include a running example using any randomized trial; I have used both the 1948 Medical Research Council trial of streptomycin for pulmonary tuberculosis ("Streptomycin Treatment of Pulmonary Tuberculosis" 1948) and the Moderna COVID-19 vaccine trial (Baden et al. 2021). Particular emphasis is placed on the statistical and ethical justifications for randomization. Mathematically, quantitative analysis of potential bias sources can be introduced, as well as the variance induced by randomization and its relation to sample size. Formal sample size and power calculations can be conducted here, or their ideas shown through computational tools and Monte Carlo simulation.

For the case studies, I have used two studies of semaglutide, the GLP-1 receptor agonist used for diabetes and obesity treatment (Lincoff et al. 2023; Wilding et al. 2021). Both trials have fairly straightforward randomization, but introduce complexity through details of the endpoints chosen, the inclusion/exclusion criteria, and the regimen and duration of follow-up. Possible discussion questions are shown in Box 2.

---

Box 2. Selected discussion questions for semaglutide randomized controlled trials.
- What could cause a lack of exchangeability if the treatment and placebo groups were not randomized?
- What sources of bias and variance remain in the randomized trial?
- What is the difference in interpretation between a study with percentage change in body weight as an endpoint (Wilding et al. 2021) and one with an endpoint of cardiovascular events usually associated with obesity (Lincoff et al. 2023)?
- How do the inclusion/exclusion criteria shape the interpretation? Are the results likely to apply to other populations?
- What ethical concerns arise from running multiple studies on the same drug product? Why are there ongoing studies of this and similar products (see, e.g., May 2024; Lenharo 2024)?

---

These studies—both timely and newsworthy—are part of ongoing research into this class of drugs. Because of the attention on the topic, there are news articles describing the studies (e.g., Blum 2023) that can be discussed as well. This allows students to both see how professional science communicators interpret the studies and critique this evaluation. Both are useful for improving the students' own statistical communication skills. The regimented reporting of clinical trials also provides an advantage for students new to reading scientific literature. For example, each study's Table 1 describes the characteristics of each treatment group, which can also make concrete some of the misconceptions around randomization (Senn 2013).

The discussion also foreshadows two key topics covered later in the course. First, the ethics of randomization can be discussed with the multiplicity of studies and with the ethics around the medicalization of obesity and crowding-out of other indications (see, e.g., Boero 2007; Oswald 2024). Second, the strict inclusion/exclusion criteria pose a limitation and create a tradeoff between internal and external validity; this will be examined from the other side when observational studies are covered, with large nutrition cohort studies discussed (Liu et al. 2022; Willett 1993).

*Quasi-experimental Designs: Economics and Policy Evaluation*

Early iterations of this course attracted a lot of students with interest in economics, political science, and public policy. Because of this, I added quasi-experiments as a topic covered in the observational studies unit. These have become particularly notable because of the Nobel Prize in Economic Sciences awarded to pioneers in quasi-experimental methods in empirical economics in 2021 (The Royal Swedish Academy of Sciences 2021) and broader debates over the appropriate types of evidence in policy analysis (see, e.g., Turner 2023). These studies also engender challenges in communicating statistical uncertainty: the assessment of causal assumptions, a lack of replication, and confusion engendered by the phrases "natural experiment" and "quasi-experiment."

I focus especially on difference-in-differences (DID) and synthetic control methods (SCM), which provide two different approaches to generating exchangeability compared to cohort studies; see, e.g., Cunningham (2021) and Huntington-Klein (2022) for textbooks covering these topics. Historical connections can be drawn to mid-1800s public health research (Caniglia and Murray 2020; Coleman 2018), demonstrating some of the intuitive appeal of this design. In addition, students can easily engage with foundational papers in the field: Card and Krueger (1994) for DID and both Abadie and Gardeazabal (2003) and Abadie et al. (2010) for SCM. Assigning two of these as the case studies demonstrates to students that methodological and study design development is ongoing, not a field consigned to the past. Selected discussion questions developed for Card and Krueger (1994) and Abadie et al. (2010) are given in Box 3. Other examples can be found in a wide array of fields. For example, having many students interested in sports analytics, I have often used evaluation of sports rule changes as running examples to introduce the methods (e.g., Kennedy-Shaffer 2022, 2024a; Sharma 2024).

---

Box 3. Selected discussion questions for quasi-experimental designs.
- Explain and evaluate—in the specific context of these studies—the assumptions required for causal inference from DID and SCM designs.
- Explain the estimand targeted in each study.
- What types of inference and sensitivity analysis are used in these studies? How do they differ from sampling-based inference?
- What other populations, settings, time frames, or policies might you want to generalize these results to? What are the dangers of this extrapolation?
- How would you describe the tradeoff between internal and external validity from using these designs to a policymaker with limited statistical background? Would the implications be different for a social scientist than for a policymaker?
- Is "natural experiment" a useful term to use for these studies? Why or why not? Compare these to randomized policies, such as those in Gay (2012) and Baicker et al. (2013).

---

These studies raise important questions about estimands, specifically the difference between a sample from a larger population compared to an analysis of all relevant units. This brings different frameworks for inference (such as randomization-based inference) into the conversation and allows disciplinary differences to be highlighted, as well as how data analysis for the ongoing process of scientific research may differ from that for specific policy or program evaluation needs. At the same time, the discussion demonstrates the value of having multiple disciplinary perspectives and being able to employ statistical thinking to critically evaluate methods used in applied fields. It also challenges the strict distinction between observational and experimental evidence, while reinforcing the value of clearly communicating the design. As these analyses become more common in formal and informal research and data analysis, and similar analyses seek to draw conclusions, it is useful for students to have a framework to evaluate assumptions and to practice interpreting and communicating results and limitations.

**Role in the Undergraduate Curriculum**

By covering topics across statistical domains and engaging with a variety of application areas, this course serves many purposes in the undergraduate curriculum. In a liberal arts setting, like the one in which the course was developed, the course allows students to be exposed to survey sampling, DoE topics, and causal inference, without the staffing burden of offering separate courses on each of these topics. It covers most of the topics laid out under "Design of Studies" in the Curriculum Guidelines topics developed by the American Statistical Association's Undergraduate Guidelines Workgroup (2014). In addition, however, it engages several other topics and domains from the Guidelines: developing more advanced understanding of point and interval estimation and hypothesis testing, reiterating the role of statistical models in data analysis, demonstrating uses of simulation and Monte Carlo methods, and using mathematical

foundations—including single- and multi-variable calculus and probability—to assess study designs. More fundamentally, the course focuses on "Statistical Practice" (including communication and collaboration), "Problem Solving" (specifically, the scientific method and the statistical problem-solving cycle and the connections between the two), and "Discipline-Specific Knowledge" across a variety of fields (American Statistical Association Undergraduate Guidelines Workgroup 2014). In larger institutions, this course can be a survey that bridges to more advanced courses in these various areas.

The course also provides an attractive offering in a minor or concentration. It can be "particularly attractive to students from certain majors", as Cannon et al. (2002) recommended for minor courses. Social sciences and life sciences and pre-health students in particular were attracted to the course; in institutions with larger populations of agriculture and engineering students, depth in traditional DoE topics may be more important instead (Vazquez and Xuan 2024). Importantly, the course is not designed primarily for students pursuing careers in statistics, but to improve statistical literacy and understanding of statistical uses that have a major impact on society (Brearley et al. 2023; Lerner and Gelman 2024; Utts 2003).

The course is not a substitute for full DoE or causal inference courses, however. It may build on the introduction to these topics that are included in introductory or intermediate methods courses (Cummiskey et al. 2020; see, e.g., Fillebrown 1994) or provide a statistical complement to more domain-specific courses in these areas (Antony and Capon 1998; see, e.g., Deming and Morgan 1983; Hiebert 2007; Zolman 1999). By providing a taste of both the domain-specific application of statistical designs and the advanced statistical theory needed for study design, the course invites students to deeper education on the aspects in which they are most interested. One recent student in the author's course noted that it "definitely improved my understanding of research studies and makes me want to continue learning more about them" ("Course Evaluations" 2023; "End of Course Survey" 2024; see Appendix B for full survey results.)

This aligns with several of the key suggestions for modernizing the undergraduate statistics curriculum (American Statistical Association Undergraduate Guidelines Workgroup 2014). Horton and Hardin (2015, p. 259) comment that statistics majors should develop "general problem solving skills to use data to make sense of the world" with an emphasis on the necessary communication skills. It serves similar goals as the DoE course proposed by Blades et al. (2015) to be the second course in the undergraduate sequence, emphasizing the problem-solving cycle and real applications, and can prepare students for a statistical capstone or consulting project (Smucker and Bailer 2015). Indeed, one student noted that it "really helped me put some of the pieces together from previous stats courses that I've taken" ("Course Evaluations" 2023; "End of Course Survey" 2024).

The flexibility of the course and its assignments allow it to be taught at various levels of mathematical understanding and computational background. Depending on the curriculum needs, it can either serve as a key nonmathematical statistics course with few pre-requisites (Higgins 1999) or as a course that connects mathematical and conceptual understanding (Green and Blankenship 2015) by using mathematics to prove properties of study designs, find optimal designs, weigh trade-offs, etc. If placed in a curriculum with computational pre-requisites, simulation can be used as a "computational engine" instead of mathematical theory (Cobb 2015). Study designs can be compared through simulation to determine operating characteristics, with an emphasis on the conceptual interpretation of the assumptions in the Monte Carlo model.

Finally, the course can both embrace big data and equip students with a statistical framework to critically consider its use. Infused with the history and social context of statistics, students can learn important lessons for the appropriate use of data science and artificial intelligence (Láinez-Moreno et al. 2024) and the dangers of over-reliance on quantification and statistical procedures that cannot be divorced from their historical context (Kennedy-Shaffer 2024b). Gaining a framework for scientific and quantitative understanding helps prevent epistemological illusions that can arise with AI (Messeri and Crockett 2024). Indeed, several writers have noted the increasing importance of study design and principled data collection for making meaning in the era of big data (see, e.g., Anderson-Cook and Lu 2023; Editorial 2024; Garrett 2024). In this way, the course prepares students to engage responsibly with modern data science tools and reinforces the importance of statistical thinking and literacy.

**Adapting the Course: Challenges and Opportunities**

The course is not without its challenges, particularly adapting it to fit the needs of both instructor and students. For one, the case studies require engagement with subject matter from a wide range of disciplines. While this helps ensure students find engaging reading at least some of the time, it can be daunting for students—and even more so for instructors—to discuss papers from fields with which they have limited familiarity. Acknowledging sites of limited knowledge builds trust with students and can highlight the importance for statisticians of engaging with subject-matter experts, literature, and methods with humility. More generally, building an inclusive and open environment for discussion is challenging in a statistics classroom and requires dedicated work by the instructor. Relevant examples and suggestions of approaches to foster this environment have been given for discussing ethics (Baumer et al. 2022; Elliott et al. 2018; Raman et al. 2023), social justice applications (Lesser 2007), and the history of statistics (Kennedy-Shaffer 2024b; Kent and Lorenat 2025). The variety, while challenging, has key pedagogical benefits; one student noted: "I especially enjoyed learning about different statistical experiments in the wide variety of fields. I feel like my perception of statistics and analyzing trials has changed because of this course" ("Course Evaluations" 2023; "End of Course Survey" 2024).

Relatedly, no single book that I am aware of covers the full suite of topics for this course. I have often used for my own reference and as optional texts for the students a mix of texts on survey sampling (e.g., Lohr 2022), experimental design (e.g., Cochran and Cox 1992; Oehlert 2000), and observational studies or causal inference (e.g., Cunningham 2021; Hernán and Robins 2024; Huntington-Klein 2022). I have also created a glossary for terms and important mathematical concepts used in the course (available upon request). While an additional challenge for instructor and student, this again opens possibilities by reducing constraints on the syllabus.

Integrating the mathematical, computational, and conceptual material also poses a challenge. While several ways of approaching this are feasible, instructors should be wary not to force everything into one course. This can be informed by the pre-requisites of the course or where it fits in the curriculum, ensuring that it complements rather than duplicates material and learning objectives in other courses. Clarifying these decisions to students is also crucial for building trust and setting expectations; in my experience, some students can be surprised by the lower computational content of the course if it is not laid out at the start of the semester. It will always be challenging, however. One student remarked: "This course was really well put together. I enjoyed the hybrid organization of discussion of material followed by practical queries of actual studies." Another in the same class commented: "I would have liked to connect the homework better with class material." Several student reviews that semester fell into each bucket, an improvement from early iterations of the course which had more of the latter comments on disjointedness between course components. Finding the right balance of conceptual discussion and computational/mathematical work will depend on the specific learning objectives, the student population and place in the curriculum, and adaptation over time.

These challenges provide opportunities as well: opportunities to bring different topics and fields into the statistics curriculum, highlight the role of statistics in science and society at large, and adapt the course to the preferences and interests of students. The active learning encouraged by discussion of the case studies also provides an avenue for engagement not often prevalent in mathematics classrooms (one student, for example, found the "lecture and discussion split-style was a nice break from some of my other very lecture heavy classes"). Brearley, Rott, and Le (2023) describe some of these benefits, albeit at the professional school level. They also highlight how modules and lessons from this course may advance statistical service education at the graduate and professional levels.

**Conclusion**

As the statistics curriculum continues to grapple with the rise of big data and artificial intelligence, engaging with core statistical principles and building statistical literacy is more important than ever. Students need to learn and practice not only the latest machine learning methods, but how to think with data, choose how and why to collect data, and communicate statistical results—and their interpretations and limitations—clearly to diverse audiences. All of these skills are crucial parts of the curriculum and can be integrated into many courses, but a course devoted to study design can put them front and center. Students engage deeply with the world around them and the roles that data and statistics should—and should not—play in policy, science, and personal decision-making.

Acknowledgements: The author wishes to thank all of the past students of MATH 348 at Vassar College. Declaration of Interests: The author reports there are no competing interests to declare.

Data Availability: The corresponding author is happy to make available any other course materials upon request.

**Supplementary Material for:**

**"An Undergraduate Course on the Statistical Principles of Research Study Design"**

Lee Kennedy-Shaffer, PhD

*Department of Biostatistics, Yale School of Public Health, New Haven, CT, USA*

Correspondence: Lee.kennedy-shaffer@yale.edu

# Appendix A: Course Materials

## A.1. Syllabus: Course Objectives, Assignments, and Grading

| Course Objectives | |
|---|---|
| **Course Overview** | Research studies are used in many fields, from economics and political science to physics, biology, and medical research. All of them share a need for statistically valid methods for study design and the analysis of results. This course covers the statistical principles and challenges behind randomized and non-randomized studies in these and other fields, highlighting the role statisticians play in the research process. Mathematical theory, examples, and simulations in R are considered in evaluating study designs. |
| **Learning Objectives** | By the end of this course, students will be able to:<br>● identify the key statistical and scientific goals of research studies;<br>● explain the benefits, drawbacks, and limitations of various study designs and features;<br>● read methods sections of scientific papers, understand the statistical concepts discussed therein, and assess the choices made in the study design;<br>● understand the role of statisticians in the design and execution of studies;<br>● communicate the course concepts to audiences with varying mathematical and statistical backgrounds;<br>● use the mathematical tools of probability and statistics to evaluate study design features; and<br>● explain how scientific goals affect statistical needs and how statistical limitations of study designs shape feasible scientific goals. |
| **Assignments and Grading** | |
| **Engagement** | You are responsible for coming to class having done any readings and ready to discuss and engage with the material (or letting me know in advance if you won't be able to attend so we can make an alternate plan). Throughout the semester, please ask questions during class and during office hours, and participate fully in group activities. |
| **Homework** | Five homeworks will be assigned roughly every other week throughout the course. These may involve reading reactions, delving more deeply into class topics, or calculating or simulating statistical properties of designs. In many cases, homework questions will go beyond the math covered in class, so be prepared to stretch your statistical thinking! |

| **Midterm Assignment** | I will post a few topics, each with an associated scientific article. You will select one of these topics and write a scientific summary of this article, placing it into context among the designs we have studied. You will also create something to explain this article to a general, non-scientist audience. |
|---|---|
| **Final Project** | In groups of 3 or 4, you will design a scientific study using the principles and designs we have discussed in class. You will conduct a short literature review on the topic and come up with a study design. You will present these results to the class and prepare a scientific paper according to the given template. |
| **Grading** | Engagement 15%<br>Homework 30%<br>Midterm Assignment 25%<br>Final Project 30% |

*A.2. Syllabus: Course Schedule*

| Week | Unit | Day 1 Topic | Day 2 Topic | Required Discussion Readings | Work Due |
|---|---|---|---|---|---|
| 1 | 1 | - | Introduction to the Study Design Process | Course Syllabus | |
| 2 | 1 | Survey Sampling; Variance | Physical Sampling: Systematic vs. Random | MacKay and Oldford 2000<br>Cochran and Cox 1957 (Ch. 1) | |
| 3 | 1 | Stratified Sampling; Bias | Political Polling: Sources of Bias | Kennedy et al. 2018<br>Gelman 2021 | HW1 |
| 4 | 2 | Introduction to RCTs and Causal Inference | Drug Trials: Exchangeability | Wilding et al. 2021<br>Lincoff et al. 2023 | |
| 5 | 2 | RCT Variance and Sample Size Calculations | Human Development: Statistical Power | James et al. 2002<br>Troller-Renfree et al. 2022 | HW2 |
| 6 | 2 | RCT Ethics and Feasibility | Social Policies: Consistency and Estimands | Bertrand and Mullainathan 2004<br>Gay 2012 | |
| 7 | 3 | Stratification and Matching | Political Access: Bias-Variance Tradeoff Stratification and Matching | Kalla and Broockman 2016 | Midterm Project |
| 8 | 3 | Multi-Arm/Factorial Designs | Agriculture: Statistical Efficiency | Cochran and Cox 1950, Ch. 5<br>Schneider et al. 2017 | HW3<br>Final Project Groups |
| 9 | 3 | Cluster Randomized Trials | COVID-19: Estimands and Variance | Abaluck et al. 2022<br>Mitjà et al. 2021 | |
| 10 | 4 | Observational Studies | Nutrition: Exchangeability and Consistency | Willett et al. 1993<br>Curtis et al. 2016<br>D'Agostino McGowan et al. 2024 | HW4<br>Final Project Topics |
| 11 | 4 | Quasi-Experimental Designs | Policy Analysis: Internal and External Validity | Card and Krueger 1994<br>Abadie et al. 2010 | |
| 12 | 4 | Case-Control Studies | Law and Study Design: Communication and Certainty | Viewing: *Erin Brockovich* 2000<br>Heath 2013<br>Steenland et al. 2014 | HW5 |
| 13 | - | Conclusion | Final Presentations I | Stevenson 2023 | Final Presentations |
| 14 | - | Final Presentations II | - | - | Final Papers |

## A.3. Syllabus: Course Topics and Reading List

*Note: Optional/supplementary readings are denoted by \*.*

**Unit 1: Survey Sampling**

*Week 2: Variance*
- Cochran WG, Cox GM. *Experimental Designs*. 2nd edn. New York: John Wiley & Sons, 1957.
- MacKay RJ, Oldford RW. Scientific method, statistical method and the speed of light. *Statistical Science*. 2000; 15(3): 254–278. https://www.jstor.org/stable/2676665.
- \* Seife C. CERN's gamble shows perils, rewards of playing the odds. *Science*. 2000; 289(5488): 2260–2262. https://doi.org/10.1126/science.289.5488.2260.
- \* Trang B. Why wastewater data isn't yet a more mainstream public health tool. *STAT News*. 11 Jan. 2023.
- \* Whitaker TB. Sampling foods for mycotoxins. *Food Additives and Contaminants*. 2006; 23(1): 50–61. https://doi.org/10.1080/02652030500241587.

*Week 3: Bias*
- Kennedy C, Blumenthal M, Clement S, et al. An evaluation of the 2016 election polls in the United States. *Public Opinion Quarterly*. 2018; 82(1): 1–33. https://doi.org/10.1093/poq/nfx047.
- Gelman A. Failure and success in political polling and election forecasting. *Statistics and Public Policy*. 2021; 8(1): 67–72. https://doi.org/10.1080/2330443X.2021.1971126.
- \* Cohn N. How one 19-year-old Illinois man is distorting national polling averages. *The New York Times*. 12 Oct. 2016. https://www.nytimes.com/2016/10/13/upshot/how-one-19-year-old-illinois-man-is-distorting-national-polling-averages.html.
- \* Graefe A, Armstrong JS, Jones RJ, Cuzán AG. Combining forecasts: an application to elections. *International Journal of Forecasting*. 2014; 30(1): 43–54. https://doi.org/10.1016/j.ijforecast.2013.02.005.

| | |
|---|---|
| **Unit 2: Introduction to Randomized Controlled Trials** | *Week 4: Exchangeability*<br>● Lincoff AM, Brown-Frandsen K, Colhoun HM, et al. Semaglutide and cardiovascular outcomes in obesity without diabetes. *New England Journal of Medicine*. 2023; 389(24): 2221–2232. http://doi.org/10.1056/NEJMoa2307563.<br>● Wilding JPH, Batterham RL, Calanna S, et al. Once-weekly semaglutide in adults with overweight or obesity. *New England Journal of Medicine*. 2021; 384(11): 989–1002. http://doi.org/10.1056/NEJMoa2307563.<br>● * Baden LR, El Sahly HM, Essink B, et al. Efficacy and safety of the mRNA-1273 SARS-CoV-2 vaccine. *New England Journal of Medicine*. 2021; 384(5): 403–416. http://doi.org/10.1056/nejmoa2035389.<br>● * Blum D. Wegovy is shown to reduce risk of heart attacks and strokes in some patients. *The New York Times*. 11 Nov. 2023. https://www.nytimes.com/2023/11/11/well/live/ozempic-wegovy-heart-disease-obesity.html.<br>● * Reis G, Silva EASM, Silva DCM, et al. Effect of early treatment with ivermectin among patients with COVID-19. *New England Journal of Medicine*. 2022; 386(18): 1721–1731. https://doi.org/10.1056/nejmoa2115869.<br>● * Marshall G, Blacklock JWS, Cameron C, et al. for the Medical Research Council Streptomycin in Tuberculosis Trials Committee. Streptomycin treatment of pulmonary tuberculosis. *British Medical Journal*. 1948; 2: 769–782. https://www.bmj.com/content/2/4582/769.<br>● * Senn S. Seven myths of randomisation in clinical trials. *Statistics in Medicine*. 2013; 32(9): 1439–1450. https://doi.org/10.1002/sim.5713.<br>● * Zimmer C, Grady D. Moderna's Covid vaccine: what you need to know. *The New York Times*. 30 Nov. 2020. https://www.nytimes.com/live/2020/moderna-covid-19-vaccine.<br><br>*Week 5: Statistical Power*<br>● James DK, Spencer CJ, Stepsis BW. Fetal learning: a prospective randomized controlled study. *Ultrasound in Obstetrics & Gynecology*. 2002; 20(5): 431–438. https://doi.org/10.1046/j.1469-0705.2002.00845.x.<br>● Troller-Renfree SV, Costanzo MA, Duncan GJ, et al. The impact of a poverty reduction intervention on infant brain activity. *Proceedings of the National Academy of Sciences USA*. 2022; 119(5): e2115649119. https://doi.org/10.1073/pnas.2115649119.<br>● * DeParle J. Cash aid to poor mothers increases brain activity in babies, study finds. *The New York Times*. 24 Jan. 2022. https://www.nytimes.com/2022/01/24/us/politics/child-tax-credit-brain-function.html. |

*Week 6: Ethics, Feasibility, Consistency*
- Bertrand M, Mullainathan S. Are Emily and Greg more employable than Lakisha and Jamal? A field experiment on labor market discrimination. *The American Economic Review*. 2004; 94(4): 991–1013. https://doi.org/10.1257/0002828042002561.
- Gay C. Moving to Opportunity: the political effects of a housing mobility experiment. *Urban Affairs Review*. 2012; 48(2): 147–179. https://doi.org/10.1177/1078087411426399.
- * Holm S. Book Review—For the Common Good. *Clinical Trials* 2024; 21(1): 136–137. https://doi.org/10.1177/17407745231193140.
- * Wilson A, Kasina F, Nduta I, Ayumbah Akallah J. When economists shut off your water. *Africa Is a Country* (web site). 2023. https://africasacountry.com/2023/11/when-economists-shut-off-your-water.
- * Jones CP. Invited commentary: 'race,' racism, and the practice of epidemiology. *American Journal of Epidemiology*. 2001; 154(4): 299–304. https://doi.org/10.1093/aje/154.4.299.

*Readings on the Tuskegee Syphilis Study:*
- * Jones JH. *Bad Blood: the Tuskegee Syphilis Experiment*. New and expanded edn. New York: Free Press, 1993.
- * Jones JH, King NMP. *Bad Blood* thirty years later: a Q&A with James H. Jones. *Journal of Law, Medicine & Ethics*. 2012; 40(4): 867–872. https://doi.org/10.1111/j.1748-720X.2012.00716.x.
- * Reverby SM, ed. *Tuskegee's Truths: Rethinking the Tuskegee Syphilis Study*. Chapel Hill, NC: University of North Carolina Press, 2012.

| **Unit 3: Advanced Topics in Randomized Controlled Trials** | *Week 7: Bias-Variance Tradeoff* |
|---|---|

*Week 7: Bias-Variance Tradeoff*
- Kalla JL, Broockman DE. Campaign contributions facilitate access to Congressional officials: a randomized field experiment. *American Journal of Political Science*. 2016; 60(3): 545–558. https://doi.org/10.1111/ajps.12180.
- * Muchnik L, Aral S, Taylor SJ. Social influence bias: a randomized experiment. *Science*. 2013; 341(6146): 647–651. https://doi.org/10.1126/science.1240466.

*Week 8: Statistical Efficiency*
- Cochran WG, Cox GM. *Experimental Designs*. Ch. 5: Factorial Experiments. 1950. New York: John Wiley & Sons. Pp. 122–153.
- Schneider M, Andres C, Trujillo G, et al. Cocoa and total system yields of organic and conventional agroforestry vs. monoculture systems in a long-term field trial in Bolivia. *Experimental Agriculture*. 2017; 53(3): 351–374. https://doi.org/10.1017/S0014479716000417.
- * Schmitz J, Hahn M, Brühl CA. Agrochemicals in field margins—an experimental field study to assess the impacts of pesticides and fertilizers on a natural plant community. *Agriculture, Ecosystems and Environment*. 2014; 193: 60–69. https://doi.org/10.1016/j.agee.2014.04.025.
- * Wood L, Welch AM. Assessment of interactive effects of elevated salinity and three pesticides on life history and behavior of southern toad (*Anaxyrus terrestris*) tadpoles. *Environmental Toxicology and Chemistry*. 2015; 34(3): 667–676. https://doi.org/10.1002/etc.2861.

*Week 9: Estimands and Effects*
- Abaluck J, Kwong LH, Styczynski A. Impact of community masking on COVID-19: a cluster-randomized trial in Bangladesh. *Science*. 2022; 375(6577): eabi9069. http://doi.org/10.1126/science.abi9069.
- Mitjà O, Corbacho-Monné M, Ubals M, et al. A cluster-randomized trial of hydroxychloroquine for prevention of COVID-19. *New England Journal of Medicine*. 2021; 384(5): 417–427. https://doi.org/10.1002/etc.2861.
- * Lipsitch M, Dean NE. Understanding COVID-19 vaccine efficacy. *Science*. 2020; 370(6518): 763–765. http://doi.org/10.1126/science.abe5938.
- * Wu A. A study in Bangladesh tripled the rate of mask-wearing. Can it help in the U.S.? *NPR Goats and Soda*. 13 Aug. 2021. https://www.npr.org/sections/goatsandsoda/2021/08/13/1027218817/what-can-the-u-s-learn-from-bangladesh-s-big-masking-experiment.

| **Unit 4: Observational Studies** | *Week 10: Exchangeability and Consistency* |

- Willett WC, Stampfer MJ, Manson JE, et al. Intake of *trans* fatty acids and risk of coronary heart disease among women. *The Lancet*. 1993; 341(8845): 581–585. https://doi.org/10.1016/0140-6736(93)90350-P.
- Curtis CJ, Clapp J, Goldstein G, Angell SY. How the Nurses' Health Study helped Americans take the trans fat out. *American Journal of Public Health*. 2016; 106(9): 1537–1539. https://doi.org/10.2105/AJPH.2016.303353.
- D'Agostino McGowan L, Gerke T, Barrett M. Causal inference is not just a statistics problem. *Journal of Statistics and Data Science Education* 2024; 32(2): 150–155. https://doi.org/10.1080/26939169.2023.2276446.
- * Liu D, Li ZH, Shen D, et al. Association of sugar-sweetened, artificially sweetened, and unsweetened coffee consumption with all-cause and cause-specific mortality. *Annals of Internal Medicine*. 2022; 175(7): 909–917. http://doi.org/10.7326/M21-2977.
- * Blum D. Coffee drinking linked to lower mortality risk, new study finds. *The New York Times*. 1 June 2022. https://www.nytimes.com/2022/06/01/well/eat/coffee-study-lower-dying-risk.html.

*Week 11: Internal and External Validity*

- Card D, Krueger AB. Minimum wages and employment: a case study of the fast-food industry in New Jersey and Pennsylvania. *American Economic Review*. 1994; 84(4): 772–793. https://www.jstor.org/stable/2118030.
- Abadie A, Diamond A, Hainmueller J. Synthetic control methods for comparative case studies: estimating the effect of California's tobacco control program. *Journal of the American Statistical Association*. 2010; 105(490): 493–505. https://doi.org/10.1198/jasa.2009.ap08746.
- * Abadie A, Gardeazabal J. The economic costs of conflict: a case study of the Basque Country. *American Economic Review*. 2003; 93(1): 113–132. http://doi.org/10.1257/000282803321455188.
- * Coleman T. Causality in the time of cholera: John Snow as a prototype for causal inference. *SSRN* [preprint]. 2019. https://doi.org/10.2139/ssrn.3262234.
- Cowger TL, et al. Lifting universal masking in schools—COVID-19 incidence among students and staff. *New England Journal of Medicine*. 2022. 387: 1935–1946. https://doi.org/10.1056/NEJMoa2211029.
- * Kennedy-Shaffer L. Baseball's natural experiment. *Significance*. 2022; 19(5): 42–45. https://doi.org/10.1111/1740-9713.01691.
- * Royal Swedish Academy of Sciences. Popular science background: natural experiments to help answer important questions. *The Prize in Economic Sciences 2021*. https://www.nobelprize.org/prizes/economic-sciences/2021/popular-information/.

- * Turner N. Impact Over Orthodoxy. *Vera Institute* (web site). 2023. https://www.vera.org/news/impact-over-orthodoxy.

*Week 12: Communication and Certainty*
- Soderbergh S. (Director). *Erin Brockovich* [Film]. Universal City, CA: Universal Studios, 2000.
- Heath D. Cancer-cluster study seeking to debunk "Erin Brockovich" has glaring weaknesses. *The Center for Public Integrity*. 3 June 2013. https://publicintegrity.org/environment/cancer-cluster-study-seeking-to-debunk-erin-brockovich-has-glaring-weaknesses/.
- Steenland K, Savitz DA, Fletcher T. Class action lawsuits: can they advance epidemiologic research? *Epidemiology*. 2014; 25(2): 167–169. http://doi.org/10.1097/EDE.0000000000000067.

*Week 13: Conclusion: Where Do We Go From Here?*
- * Messeri L, Crockett MJ. Artificial intelligence and illusions of understanding in scientific research. *Nature*. 2024; 627: 49–58. https://doi.org/10.1038/s41586-024-07146-0.
- * Stevenson MT. Cause, effect, and the structure of the social world. *Boston University Law Review*. 2023; 103: 2001–2047. http://doi.org/10.2139/ssrn.4445710.

*A.4. Study Reading Guide*

| Study Basics & Goals ||
|---|---|
| Citation Information | |
| Scientific Question of Interest | |
| Population of Interest | |
| Overall Study Type | |
| Target Audience | |

| Study Implementation ||
|---|---|
| Primary Outcome | |
| Primary Exposures/Treatments | |
| Covariates, Stratification Factors, etc. | |
| Sampling & Randomization Approach (including Sample Size) | |
| Statistical Question & Estimand | |
| Statistical Analysis Method/Estimator | |

| *Study Results* | |
|---|---|
| Estimate & Inference Values | |
| Study Conclusion & Reporting | |
| Any Secondary Analyses (Outcomes, Exposures, Etc.) | |
| Ethical & Logistical Constraints | |

| *Your Opinions* | |
|---|---|
| How well do the estimator, estimand, and scientific question of interest match? | |
| Do the treatment arms and outcomes reflect the most important question? Do they pose a risk of measurement error and/or missing data? | |
| How did ethical, logistical, and statistical features lead to trade-offs in the design? | |
| Are there any design decisions you would make differently? | |
| Are the results reported appropriately, responsibly, and in a way accessible to the target audience? | |
| Is this study convincing to you? Why or why not? | |

**Appendix B: Course Evaluations**

Note: The quotations below are the full set of responses to the open-ended final question on the anonymous Course Evaluations (May 2023) and End of Course Survey (May 2024) administered by Vassar College at the conclusion of the author's course in the spring semesters of 2023 and 2024. In total, 34 students provided a response to this question; four are omitted entirely as they are only directed to the instructor and not generally applicable to the course. All responses were collected prior to students receiving their final grades and were unavailable to the instructor until after grades were submitted. They are provided in no particular order. The responses are redacted to ensure student anonymity and comments solely pertaining to the instructor and not the course itself are omitted. All omissions are indicated by "…".

1. This was an amazing course! I loved learning more about statistical designs and this really helped me put some of the pieces together from previous stats courses that I've taken. Professor Kennedy-Shaffer is an amazing professor; he explains tough concepts very well and is very approachable. I'm really grateful for the chance to take this class at Vassar and learn from not only the professor but everyone else in the class. The lecture and discussion split-style was a nice break from some of my other very lecture heavy classes, and helped me identify some other concepts I was having a hard time with. Overall, this has been one of my favorite classes I've taken at Vassar.
2. This was a very interesting course and it definitely improved my understanding of research studies and makes me want to continue learning more about them.
3. This class was very interesting and I think it has helped me a lot as I want to go into research design and handling data in the future.
4. The course greatly expanded my knowledge of study design
5. Prof. Kennedy Shaffer was an excellent teacher who facilitated discussion and lectures very well. His homework really further my understanding of the material and he was a great help in class and during office hours. Taking this class with him definitely increased my passion for Statistics. …

6. … This course was an extremely effective mix of technical mathematics and real world applications that felt extremely relevant and interesting.
7. … I really enjoyed this course and was very helpful for understanding the data analysis and methods required for social science generally. The homework was hard, I think if I wasn't getting support from … I would have really struggled. I am happy with the little bit of R I picked up from this course too. The group project was the hardest part of this course though, my group and I struggled to work together and that made things really difficult. It did not feel like people were very motivated to work on the project at times which was hard. …
8. I really enjoyed my time in this class and appreciate Professor Kennedy-Shaffer for his efforts inside and outside the classroom in meeting the needs of his students. …
9. I loved the course. It was one of the most useful and interesting classes I have ever taken. I especially enjoyed learning about different statistical experiments in the wide variety of fields. I feel like my perception of statistics and analyzing trials has changed because of this course. Professor Kennedy-Shaffer was very accessible and helpful the entire way through. However, I felt like the homework deviated from the purpose of the class, because it was more theoretical. The homework was very challenging at first but as the semester went on I enjoyed it more and more and saw why it was useful for understanding the class. This class was truly amazing and really made me think in a statistical manner.
10. I had an overall good experience taking this course because the work load was very fair and the professor was helpful.
11. I thought this course was really interesting, and it definitely made me think about study design and statistics more deeply in my life and in my research. While the lectures were a bit dry, the content was interesting and I enjoyed learning about it. Homeworks were extremely long.
12. This was a very interesting course that put me in a good place before graduating. I would have liked to connect the homework better with class material, but I still learned a lot. The group discussions were really effective, although it would be good to incorporate the readings in the homework.

13. This course was really well put together. I enjoyed the hybrid organization of discussion of material followed by practical queries of actual studies. I learned a lot about the tradeoffs of designs and understand a lot more about the different types of designs available and when they are best applicable.
14. The class was very engaging. I enjoyed learning about the topic and doing the readings. I think I would have liked more resources and practice problems for the statistics and probability stuff that we saw on the homework. But overall it was a great class
15. Strengths: Enjoyed how there is an emphasis on contextualizing everything we learned in the broad statistics and research space. Enjoyed also what are the limitations of the types of studies. Homework was also very detailed … Weaknesses/changes: I hoped we could dive more into about how to communicate statistics effectively and not just point out flawed aspects of communication.
16. I would love our homework to be talked about more in depth during the weeks leading up to it. I feel as if there was a slight disconnect in the homework and the class …
17. I thought that Prof. Kennedy-Shaffer did a great job at meeting the objectives of the course! He pushed us to make connections to the statistical principles that we covered throughout the course to better assess certain research designs. The papers chosen covered a wide range of disciplines that added variety to the discussions, which I really appreciated. Overall, I learned a great deal about research design that I will take with me in any discipline that I will go into!
18. I think that the course and instructor did exactly what it was meant to do and I would recommend to others interested in the subject.
19. I think this was a good class overall! I think that there may be a benefit from tying the homeworks into the class discussions further, since they often feel somewhat disjointed, but I also don't think was a particularly detrimental issue.
20. I really liked the mixture of theory-based study design methods and the statistical trade offs of various design methods in combination with reading published studies through a critical lense. Homeworks I feel were quite challenging and time consuming and attending office hours was necessary, though I definitely think they helped my overall understanding.

21. I really enjoyed the lecture part of the class and thought that we examined some interesting topics and explored them in ways I had not thought about before. I felt that the homework seemed kind of unrelated to the lecture. I understood the homework and why it was assigned, but I wish the lecture and homework had related better to each other.
22. I really enjoyed this course. ... My preference in this course would have been for more focus on RCTs and quasi-experimental designs and less on sampling. But--obviously--professors should not design courses with only me in mind, and I think the way you balanced different goals makes sense. …
23. I liked the premise of this course, and I really liked the structure of one day of teaching and one day of articles every week. I wish there was more time devoted to applying the concepts before the homeworks as I found myself spending more time on these that most other classes, and I still didn't understand some of the questions and concepts. I liked the midterm and final project structures and I think those should be kept the same.
24. I enjoyed the class. Thank you
25. Homeworks were sometimes difficult, …
26. His homework assignments were definitely very difficult in comparison to what we would learn/talk about in class. I do enjoy the structure of the class where Monday would be lecture oriented while Wednesday would be discussion influenced based on reading(s) we would have to do. I simply wish that the readings we had to do related more to the homework assignments he gave.
27. Great organization and readings that tied into class. Homework problems were more difficult than expected due to little examples with numbers in class.
28. Excellent course to conclude my math education
29. Although the class topics were really interesting, I felt as if the homework was unrelated / really number heavy when the class discussions were very not. So I learned a lot through the classes about the different kind of studies, but felt as if the homework was a chore rather than something to supplement my learning in class